# On electron imaging of small molecules, Part II: tomographic reconstruction from defocus series

T.E. Gureyev[*,1,2,3,4], H.M. Quiney[1], A. Kozlov[1] and L.J. Allen[1]

[1)] *ARC Centre in Advanced Molecular Imaging, School of Physics, The University of Melbourne, Parkville 3010, Australia*

[2)] *School of Physics and Astronomy, Monash University, Clayton 3800, Australia*

[3)] *School of Science and Technology, University of New England, Armidale 2351, Australia*

[4)] *Faculty of Health Science, The University of Sydney, Sydney 2006, Australia*

**Abstract**

A practical method utilising three-dimensional image pattern matching is proposed which, in principle, is capable of unambiguous determination of the types and positions of atoms in small molecules from defocus series collected at only a few different angular orientations of the molecule. Numerical tests of the method are presented using multislice calculations of defocus series of small biological molecules. The proposed technique can in future gain from Bayesian or machine-learning approaches and is likely to be useful in cryogenic electron microscopy.

**1. Introduction**

It is known that in transmission electron microscopy (TEM) of non-crystalline and sufficiently thin specimens "diffraction tomography" (DT) [1-3] can be used to properly account for the Fresnel diffraction in the sample and correctly reconstruct the three-dimensional (3D) distribution of the complex refractive index, or, equivalently, the spatial distribution of the electrostatic potential in electron imaging. The DT approach can be based on the first Rytov or first Born approximation, instead of the projection approximation utilized in the conventional CT. The use of DT for 3D reconstruction of small molecules is discussed in detail in the first part of this work [4]. Here we develop and test a different method for reconstruction of the 3D structure of small molecules from TEM defocus series collected at a fixed rotational orientation of the molecule or a small number of different orientations.

While reminiscent in its foundations of the Big Bang Tomography technique developed by D. Van Dyck and co-authors for materials science applications [5-7], the method presented in this paper is primarily aimed at a different class of applications and utilizes a very different reconstruction approach. Namely, we suggest the use of three-dimensional image pattern matching between the simulated defocus series of single atoms and experimentally collected defocus series of unknown molecules. We show theoretically that in the case of a sufficiently sparse structure (such as a small biological molecule) the types and 3D locations of individual atoms can, in principle, be unambiguously determined from TEM defocus series collected at a single angular orientation of the structure or at a few such orientations. In essence, our theoretical investigation shows that the assumed sparsity and sharp localization of individual objects (atoms) in a 3D structure (molecule) allows one to reliably recover the information about the types and 3D location of atoms from the defocused series images, even though the latter data is limited to a 2D Ewald sphere. This is confirmed by numerical simulations presented below, which have been carried out for small biological molecules.

[*] Corresponding author. *E-mail address*: timur.gureyev@unimelb.edu.au (T.E. Gureyev).

Our current implementation of the proposed tomographic method is deterministic and is based on the minimization of the absolute pixel-wise difference between a shifted 3D template which corresponds to the defocus series of a single atom, and a 3D image corresponding to the defocus series of the unknown structure. However, it seems obvious that Bayesian or Deep Machine Learning / Artificial Intelligence (AI) approaches are likely to be particularly suitable for this method, especially when working with noisy data and uncertainties about some experimental parameters, such as the orientations of the molecule or the defocus distances [8]. Note that the problem of atomic structure determination from defocus series is reminiscent to that of facial recognition in a moving crowd of people, the latter technique being commonly implemented with the use of AI. In a practical setting typical, for example, for cryo-EM, the PMT method can be combined with the existing algorithms for classification of the orientations of the molecules in TEM images into a single Bayesian or AI based approach. The two parts of such a combined algorithm can naturally complement each other and allow the user to fully utilize the information content available in experimental images, based on the physics of TEM image formation.

## 2. Defocused images of sparse atomic structures

Consider an imaging setup with a monochromatic plane wave $I_{in}^{1/2} \exp(i2\pi kz)$ illuminating a weakly scattering object, where $k = 1/\lambda$ is the wave number, $I_{in} = const$ is the intensity of the wave and $\mathbf{r} \equiv (x, y, z)$ is the Cartesian coordinate system in 3D space. The complex amplitude $U(\mathbf{r})$ of the wave inside the object satisfies the stationary wave equation: $\nabla^2 U(\mathbf{r}) + 4\pi^2 n^2(\mathbf{r}) k^2 U(\mathbf{r}) = 0$, where $n(\mathbf{r})$ is the refractive index. In the case of electron microscopy, one has $n(\mathbf{r}) \cong 1 + V(\mathbf{r})/(2E)$, where $V(\mathbf{r}) \geq 0$ is the electrostatic potential, $E$ is the accelerating voltage and $V(\mathbf{r})/(2E)$ is typically much less than 1 (see e.g. [9]). We consider the problem of reconstruction of the 3D distribution of the electrostatic potential from intensity of transmitted waves measured at some distances from the object (defocused images), for a single or a small number of different rotational positions of the object.

Consider the situation where the object is thicker than the depth of focus, but is so weakly scattering, that the multiple scattering can be safely ignored and the first Born approximation can be applied. In this case, the change of the incident wave upon propagation through the object can be neglected and the intensity of the projection image collected at a position *z* downstream from the object along the optical axis can be expressed as an incoherent sum of the primary beam intensity and the intensities scattered by each atom independently when illuminated by the unperturbed incident plane wave [4,10]. For the convenience of the reader, we reproduce the corresponding results from [4] in the remainder of this paragraph. In particular, the 2D Fourier transform of the intensity of a defocused image can be written as

$$(\mathbf{F}_2 I)(\mathbf{q}_\perp, z) / I_{in} = \delta(\mathbf{q}_\perp) + [2\pi/(\lambda E)] \int \sin[\pi\lambda(z-z')q_\perp^2](\mathbf{F}_2 V)(\mathbf{q}_\perp, z') dz', \qquad (1)$$

where $(\mathbf{F}_2 f)(\mathbf{q}_\perp) \equiv \iint \exp[-i2\pi \mathbf{q}_\perp \mathbf{r}_\perp] f(\mathbf{r}_\perp) d\mathbf{r}_\perp$ is the 2D Fourier transform, $\mathbf{q}_\perp \equiv (q_x, q_y)$ and $q_\perp \equiv |\mathbf{q}_\perp|$. Expressing the sine function under the integral sign in Eq. (1) via a difference of two complex exponents and introducing the contrast function, $K(\mathbf{r}_\perp, z) \equiv 1 - I(\mathbf{r}_\perp, z)/I_{in}$, we arrive at

$$\begin{aligned}&(\mathbf{F}_2 K)(\mathbf{q}_\perp, z) \\ &= [i\pi/(\lambda E)][\exp(i\pi\lambda z q_\perp^2)(\mathbf{F}_3 V)(\mathbf{q}_\perp, (\lambda/2)q_\perp^2) - \exp(-i\pi\lambda z q_\perp^2)(\mathbf{F}_3 V)(\mathbf{q}_\perp, -(\lambda/2)q_\perp^2)],\end{aligned} \qquad (2)$$

where $(\mathbf{F}_3 V)(\mathbf{q}_\perp, q_z) \equiv \iint \exp[-i2\pi(\mathbf{r}_\perp \mathbf{q}_\perp + z q_z)] V(\mathbf{r}_\perp, z) d\mathbf{r}_\perp dz$ is the 3D Fourier transform of the electrostatic potential. This equation is reminiscent of the Fourier slice theorem which is used as a basis for object reconstruction in CT [3,4,10].

Consider now the model case where the sample has a "sparse" and sharply localized structure, i.e. the distribution of the electrostatic potential in the sample can be represented in the form

$$V(\mathbf{r}) = 2E \sum_{m=1}^{M} c_m \delta(\mathbf{r}_\perp - \mathbf{r}_\perp^{(m)}) \delta(z - z^{(m)}), \qquad (3)$$

where $\mathbf{r}^{(m)} \equiv (\mathbf{r}_\perp^{(m)}, z^{(m)})$ are the positions of individual "atoms" and $c_m$ are the "integral values" of the potential at each such position (these coefficients have dimensionality m³). Taking a 2D

Fourier transform of Eq. (3) and substituting the result into Eq. (1), we obtain

$$(\mathbf{F}_2 I)(\mathbf{q}_\perp, z) / I_{in} = \delta(\mathbf{q}_\perp) - (4\pi/\lambda) \sum_{m=1}^{M} c_m \sin[\pi\lambda(z^{(m)} - z)q_\perp^2] \exp(-i2\pi \mathbf{q}_\perp \mathbf{r}_\perp^{(m)}).$$

The 2D inverse Fourier transform of this expression is equal to

$$I(\mathbf{r}_\perp, z) / I_{in} = 1 - (4\pi/\lambda) \sum_{m=1}^{M} c_m \int d\mathbf{q}_\perp \exp[i2\pi \mathbf{q}_\perp (\mathbf{r}_\perp - \mathbf{r}_\perp^{(m)})] \sin[\pi\lambda(z^{(m)} - z)q_\perp^2]$$

$$= 1 + i(2\pi/\lambda) \sum_{m=1}^{M} c_m \int d\mathbf{q}_\perp \exp[i2\pi \mathbf{q}_\perp (\mathbf{r}_\perp - \mathbf{r}_\perp^{(m)})] \{\exp[i\pi\lambda(z^{(m)} - z)q_\perp^2] - \exp[-i\pi\lambda(z^{(m)} - z)q_\perp^2]\}.$$

The latter expression corresponds to 2D Fresnel diffraction integrals and can be easily evaluated (see Appendix A4 in [11]), with the resulting contrast function

$$K(\mathbf{r}) = \frac{4\pi}{\lambda^2} \sum_{m=1}^{M} c_m \frac{\cos\left[\frac{\pi |\mathbf{r}_\perp - \mathbf{r}_\perp^{(m)}|^2}{\lambda(z - z^{(m)})}\right]}{z - z^{(m)}}. \tag{4}$$

Note that the expression on the right-hand side of Eq. (4) is dimensionless. As will be clear from the numerical examples presented below, the general structure of this equation already correctly reflects some of the key features of typical defocus series of sparse atomic structures, including the contrast sign reversal at points $z = z^{(m)}$. However, the contrast function in Eq. (4) has singularities at these contrast reversal points, which is a direct consequence of the singularity of delta functions in Eq. (3). In order to make the model for $V(\mathbf{r})$ more realistic and get rid of the singularities, we now convolve each term on the right-hand side of Eq. (3) with a Gaussian $G^{(m)}(\mathbf{r}) \equiv g^{(m)}(x) g^{(m)}(y) g^{(m)}(z)$, where $g^{(m)}(x) \equiv (2\pi\sigma_m^2)^{-1/2} \exp[-x^2/(2\sigma_m^2)]$, etc., so that

$$V(\mathbf{r}) = 2E \sum_{m=1}^{M} c_m G_\perp^{(m)}(\mathbf{r}_\perp - \mathbf{r}_\perp^{(m)}) g^{(m)}(z - z^{(m)}), \tag{5}$$

with $G_\perp^{(m)}(\mathbf{r}_\perp) \equiv g^{(m)}(x) g^{(m)}(y)$. Because of the linearity of Eq. (1) with respect to $V(\mathbf{r})$, we can now simply convolve each term on the right-hand side of Eq. (4) with the corresponding Gaussian in order to obtain the relevant expression for the contrast function using the model of the potential from Eq. (5). With respect to integration over $z$ in these convolutions, we follow the approach used in multislice method [12] and replace the $z$-convolutions of the propagator terms $[z - z^{(m)}]^{-1} \cos[(\pi/\lambda) |\mathbf{r}_\perp - \mathbf{r}_\perp^{(m)}|^2 / (z - z^{(m)})]$ with $g^{(m)}(z)$ by the products of the value of the propagator term at point $z$ and the $z$-projected value of the potential, which is equal to 1, since $\int g^{(m)}(z) dz = 1$. The convolutions with respect to *x* and *y* can be calculated explicitly, since they correspond to Fresnel propagation by distances $\pm(z - z^{(m)})$ of the two-dimensional Gaussians:

$$K(\mathbf{r}) = \frac{1}{\lambda^2} \sum_{m=1}^{M} \frac{c_m}{\sigma_m^2 (z - z^{(m)})} \int d\mathbf{r}'_\perp \exp\left[\frac{-|\mathbf{r}'_\perp|^2}{2\sigma_m^2}\right] \exp\left[\frac{i\pi |\mathbf{r}_\perp - \mathbf{r}_\perp^{(m)} + \mathbf{r}'_\perp|^2}{\lambda(z - z^{(m)})}\right] + c.c., \qquad (6)$$

where "*c.c*" means the complex conjugate of the preceding expression. Note that the different treatment of the convolutions with respect to *z* and $\mathbf{r}_\perp$ are justified here in view of the paraxial approximation adopted in this problem (corresponding to a situation dominated by small angle scattering).

Using the known expression for paraxial propagation of Gaussian beams [13], we obtain directly from Eq. (6):

$$K(\mathbf{r}) = \frac{4}{\lambda} \sum_{m=1}^{M} \frac{c_m}{w_m^2(z)} \exp\left[\frac{-|\mathbf{r}_\perp - \mathbf{r}_\perp^{(m)}|^2}{w_m^2(z)}\right] \sin\left[\frac{\Delta \tilde{z}_m |\mathbf{r}_\perp - \mathbf{r}_\perp^{(m)}|^2}{w_m^2(z)} - \arctan(\Delta \tilde{z}_m)\right], \qquad (7)$$

where $\Delta \tilde{z}_m \equiv \lambda(z - z^{(m)}) / (2\pi\sigma_m^2)$ and $w_m^2(z) \equiv 2\sigma_m^2[1 + (\Delta \tilde{z}_m)^2]$. The main properties of the contrast function in Eq. (7) are as follows.

(i) The sine factor in each term under the sum in Eq. (7) describes the oscillatory profile of the propagated intensity. This term also makes the contrast function change its sign along the *z* coordinate at the point $z = z^{(m)}$, as a consequence of the definition of $\Delta \tilde{z}_m = \lambda(z - z^{(m)}) / (2\pi\sigma_m^2)$ and the fact that both sine and arctan functions are odd. Related to this change of sign is also the fact that the contrast produced by a given atom with index *m* is equal to zero in the plane $z = z^{(m)}$ (which is a consequence of the weak phase approximation used in the present model). If the atomic structure is sparse, the contribution of atoms with indices $m' \neq m$ to the contrast function at point $(\mathbf{r}_\perp^{(m)}, z^{(m)})$ is weak (see below), and hence the point of disappearance of the contrast can be easily located in the through-focus series and the spatial positions $\mathbf{r}^{(m)} = (\mathbf{r}_\perp^{(m)}, z^{(m)})$ of individual atoms can be located by this means. The type of the atom located at position $\mathbf{r}^{(m)}$ is determined by the value of $c_m$, in the model of Eq. (5). This value can be found from the magnitude of the change of the contrast

(ii) The exponential factor in each term under the sum in Eq. (7) determines the strong transverse localization of the contrast function around the axis $\mathbf{r}_\perp = \mathbf{r}_\perp^{(m)}$, parallel to $z$, making each term in the sum behave as a Gaussian beam with the waist radius $w_0^2 \equiv w_m^2(0) = 2\sigma_m^2$, achieved in the plane $z = z^{(m)}$. The longitudinal localization (i.e the localization along the $z$ coordinate) of the contrast function in Eq. (7) is determined primarily by the factor $w_m^{-2}(z) \equiv 0.5\sigma_m^{-2}[1+(\Delta \tilde{z}_m)^2]^{-1} = 2\pi^2 \sigma_m^2 / [(2\pi\sigma_m^2)^2 + \lambda^2(z-z^{(m)})^2]$. Therefore, the diffraction contrast decreases along $z$ generally in proportion to the square of the distance from the plane in which the atom is located. In this sense, the longitudinal localization of the "traces" of atoms in the defocus series is weaker than their transverse localization and the structure needs to be more sparse in the $z$ direction, compared to $(x, y)$, in order to avoid the "screening" of the defocused images of some atoms by other atoms located on the same or adjacent lines $z =$ constant.

(iii) Note that the parameters $c_m$ are not independent of the parameters $\sigma_m$, as both of them are determined by the type of a given atom. If the atomic structure in Eq. (5) is sparse, the $4M$ independent parameters, e.g. $(c_m, \mathbf{r}^{(m)})$, $m=1,...,M$, can be found from 2D images collected at defocus distances around $z = z^{(m)}$. At such defocus positions, generally only the atom located at $(\mathbf{r}_\perp^{(m)}, z^{(m)})$ will contribute non-negligibly to the contrast function around this point, since the contribution from all other terms in Eq. (7) will be strongly attenuated by the small factors $w_m^{-2}(z)\exp[-w_m^{-2}(z)|\mathbf{r}_\perp - \mathbf{r}_\perp^{(m)}|^2]$, if the sparsity condition is satisfied and the abovementioned "screening" can be avoided.

The corresponding picture in the reciprocal space can be obtained by substituting the 3D Fourier transform of Eq. (5) into Eq. (2) and then taking the 1D Fourier transform with respect to $z$:

$$(\mathbf{F}_3 K)(\mathbf{q}) = (i2\pi/\lambda)\sum_{m=1}^{M} c_m \exp\{-2\pi^2 \sigma_m^2 q_\perp^2 [1 + (\lambda/2)^2 q_\perp^2]\}\exp(-i2\pi\mathbf{r}_\perp^{(m)}\mathbf{q}_\perp)$$
$$\times \{\exp(-i\pi\lambda z^{(m)} q_\perp^2)\delta[q_z - (\lambda/2)q_\perp^2] - \exp(i\pi\lambda z^{(m)} q_\perp^2)\delta[q_z + (\lambda/2)q_\perp^2]\}. \tag{8}$$

Note that Eq. (8) represents a 3D Fourier transform of intensity registered in the Fresnel region, rather than the intensity of a complex wave in the reciprocal space, which can be registered, for example, in the far-field diffraction regime. This equation shows that, at a fixed angular orientation of the molecule, the non-zero contrast is limited to the paraboloids $q_z = \pm(\lambda/2)q_\perp^2$.

For all atoms, the contrast exponentially decreases as a function of reciprocal coordinates, according to the term $\exp\{-2\pi^2\sigma_m^2 q_\perp^2[1+(\lambda/2)^2 q_\perp^2]\}$. However, since the Gaussians in Eq. (5) are supposed to be narrow, with the width of approximately 1 Å, the latter exponential terms are going to be broad in the reciprocal space, with the width of the order of 1 Å$^{-1}$. Therefore, the diffraction "signal" from different atoms in the model of Eq. (5) will be sufficiently strongly represented on the support of the delta functions in Eq. (8). The corresponding phases depend on the projection of the atom location vector $(\mathbf{r}_\perp^{(m)}, z^{(m)})$ onto the reciprocal vectors $\mathbf{q}$, according to the term $\exp\{-i2\pi[\mathbf{r}_\perp^{(m)}\mathbf{q}_\perp \mp z^{(m)}(\lambda/2)q_\perp^2]\}$. The information about the atoms' types and 3D locations can be extracted from this type of data, in principle. However, the practical method for solving the problem of 3D reconstruction of atomic positions and types that we develop in the next section is based on the real space formulation given by Eq. (7), rather than reciprocal space description of Eq. (8). Note that Eq. (7) can be obtained by the inverse 3D Fourier transform of Eq. (8), provided that we ignore the terms $(\lambda/2)^2 q_\perp^2$ in the first exponent. If we assume, as above, that the desired spatial resolution is close to 1 Å and the electron wavelength is around 0.025 Å, then we get $(\lambda/2)^2(q_{x,\max}^2 + q_{y,\max}^2) < 2\times 10^{-4}$. Therefore, in such cases, $(\lambda/2)^2 q_\perp^2 \ll 1$ and hence the approximation required for the derivation of Eq. (7) is well justified.

## 3. Pattern matching tomography

Here we describe a method for 3D reconstruction of sparse atomic structures from TEM defocus series exploiting the theoretical developments discussed in Section 2. This technique has similarities in its general approach with the 'Big Bang Tomography' method described in [4,5]. However, as the method developed in the present paper, unlike Big Bang Tomography, is substantially based on 3D image pattern matching, it is termed here Pattern Matching Tomography (PMT).

An essential feature of Eq. (7) is that the defocus contrast function $K(\mathbf{r})$ depends only on the relative differences $\mathbf{r} - \mathbf{r}^{(m)}$ between the point $\mathbf{r}$ and the atomic positions $\mathbf{r}^{(m)}$. Therefore, a uniform shift of all atomic positions by some vector $\mathbf{s}$, $\mathbf{r}^{(m)} \to \mathbf{r}^{(m)} + \mathbf{s}$, is equivalent to the shift of the contrast distribution, $K(\mathbf{r}) \to K(\mathbf{r}+\mathbf{s})$. This allows one, in principle, to find an unknown spatial position of a known "template" structure, e.g. a single atom, by spatially shifting a suitably simulated defocus series of that structure in space and finding the shift position that delivers the best match with the experimentally collected defocus series. Additionally, the sparsity assumption described above allows one to search the defocus series of a sparse atomic structure for the spatial position of each single atom, independently of other atoms in the structure.

A practical PMT reconstruction can in principle be based on just two 2D intensity images collected at close defocus distances $z_1$ and $z_2$, at a fixed rotational position of the sample. The two intensity images can be used to uniquely retrieve the phase of the transmitted wave in the plane $z_1$, e.g. by using the well-known TIE algorithm [14]. This will make the complex amplitude of the transmitted wave available in this plane, which can then be used to calculate defocused images at multiple distances, and use these images for locating the atoms and finding the unknown parameters $(c_m, \sigma_m, \mathbf{r}^{(m)})$ as described above. Such a technique can provide information about the locations and the types of atoms present in the sparse atomic structure that is being investigated. Alternatively, multiple defocus images can be collected experimentally at one or more angular orientations of the unknown structure. In both cases, the subsequent reconstruction of the 3D sample from these numerically or experimentally acquired defocus image series can be processed according to the same algorithm described below.

A 3D pattern matching problem can often be solved by means of maximising the correlation integral of the two patterns, while shifting one of the patterns by different amounts in *x*, *y*, and *z* directions with respect to the second pattern. This procedure can be effectively implemented using Fourier transforms. Note, however, that if two patterns differ by a multiplicative factor, their correlation coefficient will be equal to one, and so the patterns will be considered perfectly matched. This may, in practice, lead to false matches between defocus patterns produced by atoms of different types. We have found that, in the present case, a somewhat simpler and more robust method can be implemented on the basis of pattern subtraction, instead of the correlation. Indeed, two patterns can be considered ideally matched, if the subtraction of one pattern from the other results in a difference consisting of all zeros. In the presence of noise or other image imperfections, such a perfect match cannot be achieved, of course, and the goal of pattern matching then reduces to the minimization of the absolute pixel-wise difference between the two patterns. Accordingly, in order to find the 3D location of an atom within the unknown atomic structure, we first create a 3D "template" pattern consisting of a defocus series of a single atom of the given type in a known location, and then truncate it down to a 3D spatial volume comparable to the atom size in (*x*, *y*) planes and to the distance between the minimum and the maximum scattered intensity in the *z* direction. We then look for the 3D shift position of this template that results in the minimal absolute difference with the same size sub-volume of the 3D "test" pattern corresponding to the measured defocus series of the unknown structure. In other words, we search for the maxima $C_{n,m}^{\max} = C_n(i_{n,m}, j_{n,m}, k_{n,m}) \equiv \max\limits_{(i,j,k) \in S_{n,m}} C_n(i,j,k)$ of the "similarity coefficient"

$$C_n(i,j,k) \equiv 1 - \frac{\sum\limits_{(i',j',k') \in T_{n,i,j,k}} |I^{test}(i',j',k') - I_n^{tempate}(i+i', j+j', k+k')|}{\sum\limits_{(i',j',k') \in T_{n,i,j,k}} [I^{test}(i',j',k') + I_n^{tempate}(i+i', j+j', k+k')]}, \qquad (9)$$

as a function of discrete shifts of center positions, $(i, j, k)$, of the 3D template pattern (truncated defocus series) $I_n^{tempate}(i', j', k')$ of a single atom of type $n$, with respect to the 3D test pattern $I^{test}(i', j', k')$, corresponding to the defocused series of the unknown structure. The indices $i, j$ and $k$ correspond respectively to the coordinates $x$, $y$ and $z$, such that e.g. $x = x_{min} + j\, x_{step}$, where $x_{step}$ is the grid step size in the $x$ direction, and similarly for the other two coordinates. The sums in Eq. (9) are taken over the set $T_{n,i,j,k}$ containing all pixel indexes corresponding to the shifted template pattern. The successive maximums $C_{n,m}^{max}$, $n = 1, 2, .., N$, $m = 1, 2, ..., M_n$, where $N$ is the number of different atomic species and $M_n$ is the number of atoms of type $n$ in the structure, are evaluated over all shifts $(i, j, k)$ from the set $S_{n,m}$ of all pixel indexes in the test pattern, excluding:

(i) all pixels near the test pattern border, such that the template pattern centered at these pixels would extend outside the test pattern; and
(ii) a pre-defined vicinity of all maximums found at the previous steps, with $n' \leq n$ and $m' < m$.

It is easy to appreciate that the "similarity coefficient" defined in Eq. (9) always satisfies the inequalities $0 \leq C_n(i, j, k) \leq 1$, with $C_n(i, j, k) = 0$ when the template does not overlap at all with the test pattern and $C_n(i, j, k) = 1$ (indicating a perfect match) when $I^{test}(i', j', k') = I_n^{tempate}(i + i', j + j', k + k')$ at each pixel of the shifted template. As explained above, the template shift $(i_{n,m}, j_{n,m}, k_{n,m})$ that results in the maximal similarity coefficient, identifies the 3D position of the $m$-th atom of type $n$ in the sample. Note that it is assumed that in the case of future experiments utilising this method, while the "test" defocus series of an unknown structure (e.g. a small molecule) will be collected using a TEM instrument, the "template" defocus series for the individual constituent atoms could still be generated numerically under the conditions consistent with the relevant TEM experiment, such as e.g. the electron energy and objective aperture.

In the current implementation of the method we assume that the number of atoms of each type present in the molecule, i.e. the values $M_n$, are known *a priori*. However, we expect that it should be also possible to implement a slightly different strategy, where one would first search for a larger number of atoms of each type, say, $M'_n \geq M_n$, and then the correct number of matched atoms of each type would be found on the basis of the relevant "cut-off" values of the coefficient $C_n(i, j, k)$. This idea is based on the assumption that a given template pattern for one atom of type $n$ will have a higher value of the similarity coefficient with the defocus "traces" of atoms of the same type in the test pattern, compared to atoms of other types or regions with no atoms.

Note also that this method contains some features characteristic of super-resolution imaging in the sense that the location of the "centre" of the template pattern (which typically corresponds to the point of contrast reversal in the defocus series) can be determined by the pattern matching method to a precision that can be significantly finer than the size of the template pattern [15]. The actual accuracy of the match will be determined in practice by the contrast-to-noise ratio of the experimental data, as well as by the accuracy of a *priori* knowledge utilised in this method, i.e. by the degree of agreement between the parameters of the model used in the method and the actual parameters of the experiment. The degree of sparsity of the unknown atomic structure is an example of the latter factor (see examples in the next section).

It is easy to understand how the above method can be extended to the cases where more than one defocus series is available, each corresponding to a different angular orientation of the unknown structure. In order not to over-complicate the notation, we will not present explicitly the rather obvious modification of Eq. (9) for this case. However, we will describe the practical steps that can be taken in such a reconstruction from multiple orientations of the sample in an example below. The method can also potentially incorporate both coherent and incoherent aberrations of an electron microscope. While such aberrations can be simulated with the help of the TEMSIM code [16] that we use below, we have not yet incorporated them in the presented examples. It can be expected that these aberrations, if sufficiently small, will lead to small deterioration of performance of the PMT method in a way not dissimilar to the effects of the finite objective aperture of the electron microscope and thermal vibrations of the molecules, which are considered in the examples below.

**4. Numerical simulations**

In this section we describe numerical simulations performed to test the potential of practical reconstruction of sparse atomic structures from TEM defocus image series collected at one or more angular orientations of the structure. The code used for these simulations was based on the TEMSIM C++ source code developed by E.J. Kirkland, which is freely available on the Github site [16]. The TEMSIM programs allow one to simulate defocused TEM images using the multislice method [17]. We have compared the output of our code with another well-known software package for TEM simulations, µSTEM [18], and made sure that the results of the two programs quantitatively agree with each other. Our code can be also accessed at Github [19].

*4.1. Example 1 - aspartate molecule*

The first example uses a very small biological molecule, aspartate, which contains only 16 atoms: $C_4 H_7 N O_4$ [20]. A representation of the molecule in the orientation used in our simulations below is shown in Fig. 1. We centred the aspartate molecule within a $10 \times 10 \times 10$ Å$^3$ cube '$Q_{10}$' which was located in the positive octant of the Cartesian coordinates ($x$, $y$, $z$), with one corner at the point (0, 0, 0) and all sides parallel to the coordinate axes.

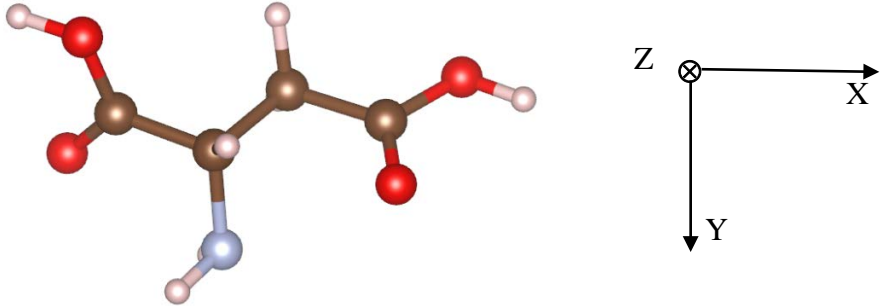

Fig. 1. Aspartate molecule as displayed by the Vesta software [21]. Oxygen atoms are red, nitrogen is blue, carbons are brown and hydrogens are pale pink. The orientation of the axes is the same as in our simulations, with $z$ axis pointing away from the viewer.

The simulation cube $Q_{10}$ was assumed to be illuminated by a plane monochromatic incident electron plane wave with an energy of 200 keV and uniform intensity equal to one, propagating along the $z$ axis. The TEMSIM code was applied for calculation of propagation of the electron wave from the 'incident' plane ($x$, $y$, 0) to the 'exit' plane ($x$, $y$, 10). The resultant transmitted complex amplitude distribution in the exit plane was then used to calculate defocused (backpropagated along the $z$ coordinate) images (intensity distributions) at 11 ($x$, $y$) planes separated by 1 Å and located between $z = 10$ Å and $z = 0$ Å, by computing the relevant Fresnel diffraction integrals. We have also produced similar defocus series for a single O, N, C or H atom located at the centre of the cube $Q_{10}$, i.e. at point (5, 5, 5). After starting with highly idealised imaging conditions, as described above, we subsequently included some realistic experimental factors, such as (i) finite aperture of the objective lens of the microscope, which was set to 30 mrad in this test; (ii) thermal vibrations of the atoms with 0.1 Å root-mean-square displacement; (iii) 1% Poisson shot noise in the detected images. The features (i) and (ii) are available in the TEMSIM code, and feature (iii) was realised using X-TRACT code [22]. Examples of an "ideal" and "realistic" defocused images are presented in Fig. 2. Note that the orientation of the aspartate molecule in this figure differs by ±90 degrees rotation around the vertical ($y$) axis with respect to Fig. 1(a) and (b) in the first part of this paper [4]: the orientation shown in Fig. 2 minimizes atomic overlap and "screening" in defocus series.

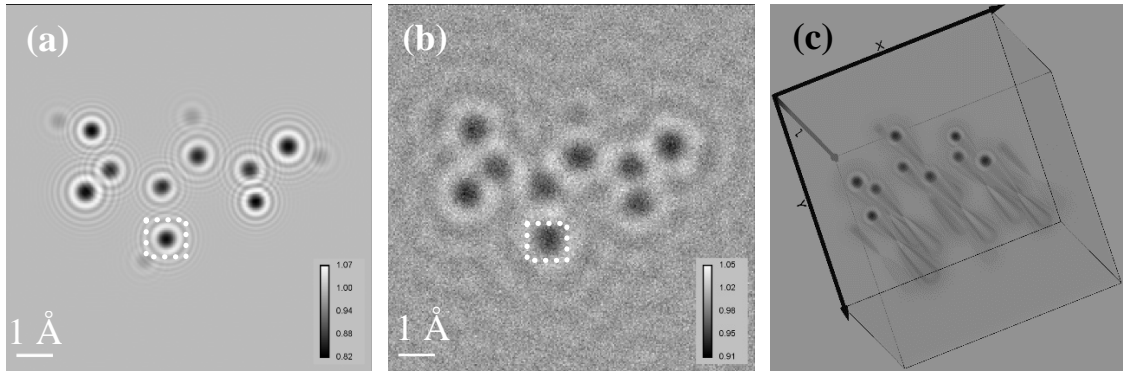

Fig. 2. Defocused images of an aspartate molecule obtained under the conditions of plane electron wave illumination (see main text for details). (a) "Ideal" image in the plane $z = 0$, i.e. at the defocus distance of $\Delta z = -10$ Å from the exit plane; (b) image in the plane $z = 0$ corresponding to objective aperture of 30 mrad, thermal vibrations with 0.1 Å root-mean-square displacement and 1% Poisson shot noise; (c) 3D rendering of defocus series in the "ideal" imaging conditions. The white dotted square in (a) and (b) indicate typical $(x, y)$ dimensions of the template pattern (see main text).

The template patterns corresponded to the single-atom defocus series with the transverse, $(x, y)$, size comparable with dimensions of a typical atom (i.e. around 1 Å$^2$). In the defocus direction $z$, the 3D template pattern had the size comparable with the distance between the first intensity extremum on each side of the point $z = z^{(m)}$ of contrast reversal at the atom position. Because the $z$ localization is weaker than the $(x, y)$ localization, as explained in Section 3, the truncated 3D template pattern typically had the shape of a parallelepiped elongated in the $z$ direction. The truncation sizes also depended on the objective aperture, because lower values of this parameter led to significant blurring of the "trace" of an atom in the defocus series, particularly, in the $z$ direction. Figure 2 shows an outline of a typical truncated 3D template pattern in the transverse plane. Figure 3 gives examples of the intensity distribution in the defocus series as a function of the defocus distance at different objective aperture values, while Figure 4 shows typical $z$ profiles of intensity in the defocused series of a single atom.

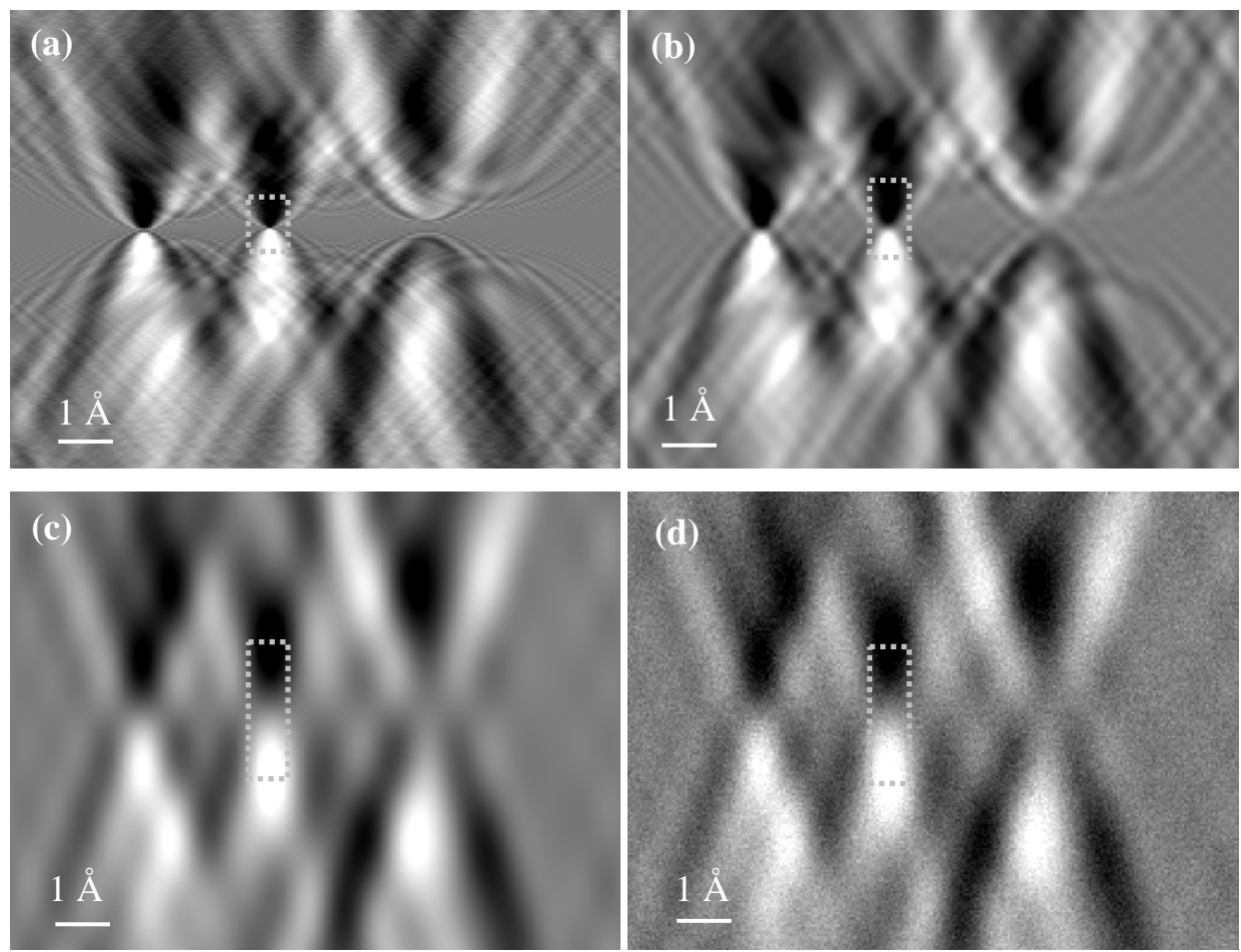

Fig. 3. Intensity distributions in (*x*, *z*) planes, at fixed value of *y* = 5, in the defocused images of an aspartate molecule: (a) "ideal" imaging conditions; (b) as in (a), but with objective aperture of 70 mrad; (c) with objective aperture of 30 mrad; (d) as in (c), but also with 1% Poisson shot noise. The grey dotted rectangle indicates typical longitudinal dimensions of a single-atom template pattern (see main text).

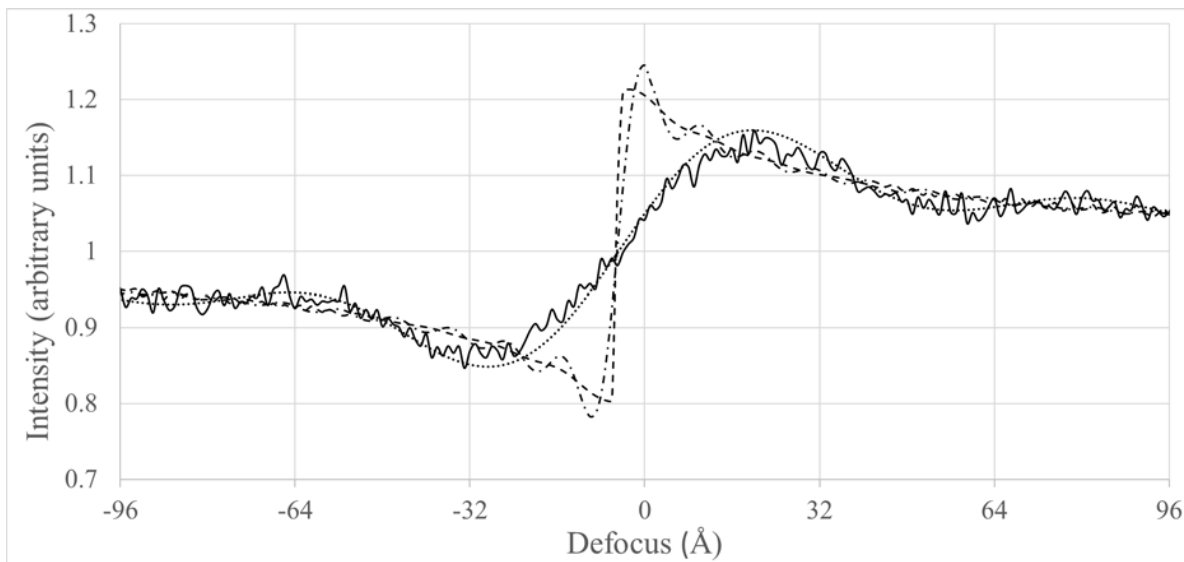

Fig. 4. Longitudinal 1D sections (along the *z* coordinate) along the central line of defocus series of a single oxygen atom: dashed line - "ideal" imaging conditions; dash-dotted line - same as "ideal", but with 70 mrad objective aperture; dotted line - 30 mrad objective aperture; solid line - 30 mrad objective aperture and 1% Poisson shot noise.

Table 1 demonstrates the results of numerical reconstruction of the aspartate molecule, in accordance with the PMT method described in Section 3, obtained from the defocus series with 11 equally spaced images between $z = 0$ Å and $z = 10$ Å, with the simulated data corresponding to objective aperture of 30 mrad, thermal vibrations with 0.1 Å root-mean-square displacement and 1% Poisson shot noise. One can see that the positions of all O, N and C atoms have been reconstructed with good (sub-Ångström) accuracy, although the three atom types have been partially mismatched (see entries number 3, 5 and 7 in Table 1). It turns out that the difference between the "signals", i.e. the difference in the image contrast, produced by these three types of atoms in the defocus series under the adopted conditions, was well below the noise level in the input data, and hence there was no realistic way to consistently distinguish these atom types without additional *a priori* information. The required *a priori* information can in principle be obtained from the known chemistry of biological molecules. However, the use of such chemical information is beyond the scope of the present paper. For the same reason, i.e. because of the very weak contrast of H atoms in the defocus series (see Fig. 2(b)), the position of these atoms could not be found. Note however, that four out of seven H atoms in the aspartate molecule could be correctly localized by the same PMT method from the defocus series under "ideal" imaging conditions (most importantly, with the infinite objective aperture). The reason for the inability of the PMT method to localize some H atoms even from these "idealized" defocused series can be understood from Fig. 1 or Fig. 2(a): it is easy to see that three out of seven hydrogen atoms are "screened" by the heavier atoms in this particular orientation of the molecule. It is likely that by using the data from additional defocus series obtained at a different rotational position of the molecule it would be possible to recover the positions of remaining H atoms. This type of procedure is demonstrated in the next example which involves a significantly larger molecule.

Table 1. Results of the PMT reconstruction of 3D positions of non-hydrogen atoms in aspartate molecule from defocus series with 30 mrad objective aperture, 0.1 Å root-mean-square displacement of thermal vibrations and 1% Poisson noise. All atomic coordinates are in Å. The similarity coefficients $C_{n,m}^{max}$ have been calculated as defined before Eq. (9). The Distance column contains the distances in Å between the original and reconstructed positions of the atoms.

|  | *Atom type* | *x* | *y* | *z* | $C_{n,m}^{max}$ | *Distance* |
|---|---|---|---|---|---|---|
| *Original 1* | O | 2.33 | 3.41 | 4.31 |  |  |
| *Reconst.1* | O | 2.34 | 3.40 | 4.00 | 0.78 | 0.31 |
| *Original 2* | O | 7.70 | 3.84 | 5.51 |  |  |
| *Reconst.2* | O | 7.70 | 3.83 | 6.00 | 0.78 | 0.49 |
| *Original 3* | O | 4.39 | 6.36 | 5.03 |  |  |
| *Reconst.3* | N | 4.38 | 6.37 | 5.00 | 0.77 | 0.03 |
| *Original 4* | O | 2.17 | 5.09 | 5.75 |  |  |
| *Reconst.4* | O | 2.15 | 5.12 | 6.00 | 0.77 | 0.25 |
| *Original 5* | N | 5.24 | 4.09 | 5.38 |  |  |
| *Reconst.5* | C | 5.20 | 4.14 | 5.00 | 0.77 | 0.38 |
| *Original 6* | C | 4.23 | 4.96 | 4.62 |  |  |
| *Reconst.6* | C | 4.26 | 4.96 | 4.00 | 0.75 | 0.62 |
| *Original 7* | C | 6.81 | 5.35 | 4.16 |  |  |
| *Reconst.7* | O | 6.80 | 5.39 | 4.00 | 0.75 | 0.16 |
| *Original 8* | C | 6.64 | 4.48 | 4.97 |  |  |
| *Reconst.8* | C | 6.60 | 4.41 | 5.00 | 0.74 | 0.08 |
| *Original 9* | C | 2.84 | 4.49 | 4.94 |  |  |
| *Reconst.9* | C | 2.93 | 4.45 | 5.00 | 0.74 | 0.12 |

*4.2. Example 2 - lasso peptide molecule*

In the second example, we attempted to reconstruct the location of atoms in the lasso peptide (LP) molecule, $S_2O_{24}N_{23}C_{95}H_{125}$, whose structure is available in the Protein Data Bank under the code 3NJW [23]. We converted the description of the structure of this molecule from the standard PDB file to the XYZ format file used by TEMSIM software [16]. As part of the conversion procedure, the molecule was centered within the cube $Q_{30}$, similarly to the case with the aspartate molecule in the previous section, but with the cube $Q_{30}$ having the side length of 30 Å, as required to accommodate the LP molecule. The TEMSIM code was applied for calculation of propagation of the plane monochromatic electron wave with $E = 200$ keV from the 'incident' plane $(x, y, 0)$ to the 'exit' plane $(x, y, 30)$, using the multislice method to calculate the propagation of the wave through the LP molecule. The resultant transmitted complex amplitude distribution in the exit plane was then used to calculate the defocused images at 31 equally spaced $(x, y)$ planes located between $z = 30$ Å and $z = 0$ Å. While we initially performed the forward simulations and PMT reconstruction under the ideal imaging conditions, here we concentrate on the most realistic of the considered cases which involved an objective aperture of 70 mrad (possible with the new generation of aberration correctors [24]), thermal vibrations with the root-mean-square deviation of 0.1 Å and 1% Poisson noise in the "registered" images. Examples of "ideal" and "realistic" defocused images of the LP molecule are presented in Fig. 5. We have also produced defocus series for a single S, O, N and C atom located at the point (15, 15, 15), using the same objective aperture and thermal vibration parameters as for the LP molecule, but not including the Poisson noise, as the latter is not relevant for the simulated "template" images.

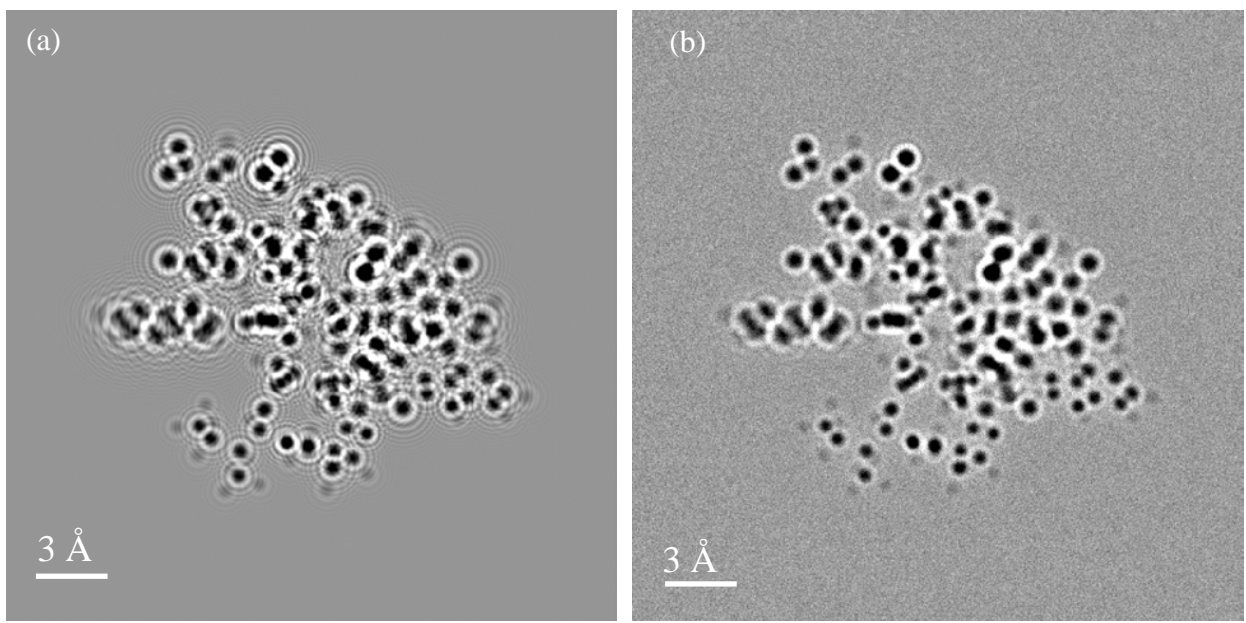

Fig. 5. Defocused images of lasso peptide 3NJW molecular structure obtained under the conditions of plane monochromatic electron wave illumination (see main text for details) in the plane $z = 0$, i.e. at the defocus distance of $\Delta z = -30$ Å from the exit plane: (a) "ideal" image; (b) image corresponding to objective aperture of 70 mrad, thermal vibrations with 0.1 Å root-mean-square displacement and 1% Poisson shot noise.

The simulated defocus series were used as input for the PMT reconstruction procedure based on Eq. (9). The details of the procedure used at this step of the simulations were very similar to the ones described above in the case of reconstruction of the aspartate molecule. This PMT reconstruction, after truncation at the level of the similarity coefficient $C_n(i_{n,m}, j_{n,m}, k_{n,m}) = 0.5$ and filtering out the "duplicate" localizations which were closer than 1 Å to each other, resulted in the correct localization of 88 out of 144 non-hydrogen atoms, with the average distance between the 3D reconstructed and the original spatial positions of the atoms equal to 0.26 Å, with the standard deviation of 0.21 Å. We conjectured that most of the missing atom locations were due to the screening effect of other atoms in the molecule, with this effect obviously being stronger in the case of the LP molecule, compared to the much smaller aspartate molecule. In order to verify this hypothesis, we also simulated two more defocus series with the same parameters as above, but with the LP molecule rotated by -20 and -45 degrees around the *y* axis, respectively. We then performed PMT reconstructions from the two additional defocus series separately, filtered the resultant atom locations by eliminating all duplicate atoms that were located closer than 1 Å to each other and truncated the filtered atom location sets at the level of similarity coefficient of 0.5. The two filtered atom location sets were then rotated back by 20 and 45 degrees, respectively, in order to bring them into the same orientation as the reconstructed set for the LP molecule in the original (0 degrees) rotational position. After that, we merged the atom locations obtained for all three angular orientations of the LP molecule into a single larger data set of atomic locations. The merged set was then again filtered for multiple atoms located closer than 1 Å to each other, leaving in the final set only one atom with the highest similarity coefficient from each such "multiplet". The logic behind this procedure was as follows. Many of the same atoms were located from each one of the three defocus series, while some other atoms were successfully localized only from two or just one of the defocus series. The conditions for localization of common atoms in different defocus series were less or more favourable for different atoms, depending on their environment (degree of screening by neighbouring atoms) in the LP molecule at a given orientation. Therefore, it is logical to observe the similarity coefficients to vary from one series to another, and to expect a higher value of the similarity coefficient to appear in a more favourable orientation for a given atom, which should also be associated with a more accurate 3D localization. Applying this logic in the PMT procedure, we were able to accurately reconstruct, from the three defocus series, the 3D positions of 120 out of 144 non-hydrogen atoms in the LP molecule. The average distance between the reconstructed and the original atom positions was equal to 0.34 Å, with the standard deviation of 0.23 Å. While the two S atoms were correctly localized and identified from each one of the three defocus series, the O, N and C atoms were partially misidentified in the individual and the final merged PMT reconstructed sets. The reason for this, just as in the case of the aspartate molecule above, was the very small difference in the diffracted intensities produced by the three types of atoms. This difference was found to be smaller than the errors in the diffracted intensity data due to the combined effect of the finite objective aperture, finite spatial resolution of the simulated input data, thermal vibration of the atoms and the shot noise in the images.

## 5. Conclusions

In the first part of this work [4], we have argued that in TEM imaging of small biological molecules or other "sparsely localized" weakly scattering structures, multiple scattering tends to have only a moderate effect and therefore can be safely ignored in the reconstruction procedures, without introducing large errors into the results. On the other hand, the in-molecule free-space propagation (Fresnel diffraction) cannot be ignored because of the extremely shallow depth of focus under the typical TEM imaging conditions. As DT represents precisely the technique which takes into account the free-space propagation, but not the multiple scattering between different atoms, it appears to be a very good match for this case, as was argued by other authors previously [10,25]. However, acquisition of images on a dense angular grid over the full $2\pi$ range, as required in DT, can often be challenging. As one way to alleviate this problem, in the present paper we have proposed a technique of pattern matching tomography, that we have called PMT for short, that fully exploits the information about the 3D atom locations as available in TEM defocus series. Such information can be naturally present in typical cryo-EM data. Instead of trying to "correct" the data for CTF variation (different defocus depths), one can fully utilize this information in the reconstruction, as proposed in our method. We have demonstrated, by means of numerical experiments utilizing multislice calculations for the simulation of TEM defocus series, that the PMT method can perform reasonably well under realistic imaging conditions. It appears that such a technique can potentially be useful, for example, in cryo-EM.

## Acknowledgements

T.E.G. and A.K. acknowledge discussions with Prof. David Paganin related to the present work. T.E.G. is also grateful to Dr. Sheridan Mayo and Mr. Darren Thompson for helpful advices. The authors wish to thank Dr. Andrew Martin for sharing his software code that was used in the course of this research.